\documentclass[layout=onecolumn, journal=jpcafh]{achemso}

\setkeys{acs}{chaptertitle = true}

\usepackage{geometry}
\usepackage{graphicx,subfigure,mciteplus}
\usepackage{amssymb, amsmath}
\usepackage[flushleft]{threeparttable}
\usepackage{epstopdf, booktabs}
\usepackage{multirow,bigdelim}
\makeatletter
\g@addto@macro\TPT@defaults{\footnotesize}
\makeatother

\usepackage{graphicx}
\usepackage{dcolumn}
\usepackage{bm}
\usepackage{longtable}
\usepackage{multirow}
\usepackage{tablefootnote}

\SectionNumbersOn

\def\deltaE{$\Delta E$}
\def\degree{$^\mathrm{o}$}

\author{Jason M. Montgomery}
\affiliation{Department of Chemistry, Biochemistry, and Physics, Florida Southern College, Lakeland, FL 33801}
\author{Ezra Alexander}
\affiliation{Department of Chemistry, Massachusetts Institute of Technology, Cambridge, MA 02139}
\author{David A. Mazziotti}
\email{damazz@uchicago.edu}
\affiliation{Department of Chemistry and The James Franck Institute,
University of Chicago, Chicago, IL 60637}

\title[Prediction of the Existence of LiCH]{Prediction of the Existence of LiCH, a Carbene-like Organometallic Molecule}

\begin{document}

%
%
%
%
%

\begin{abstract}
Carbenes comprise a well-known class of organometallic compounds consisting of a neutral, divalent carbon and two unshared electrons.  Carbenes can have singlet or triplet ground states, each giving rise to a distinct reactivity.  Methylene, CH$_2$, the parent hydride, is well-known to be bent in its triplet ground state.  Here we predict the existence of LiCH, a carbene-like organometallic molecule. Computationally, we treat the electronic structure with parametric and variational two-electron reduced density matrix (2-RDM) methods, which are capable of capturing multireference correlation typically associated with the singlet state of a diradical.  Similar to methylene, LiCH is a triplet ground state with a predicted 15.8 kcal/mol singlet-triplet gap. However, unlike methylene, LiCH is linear in both the triplet state and the lowest excited singlet state. Furthermore, the singlet state is found to exhibit strong electron correlation as a diradical.  In comparison to dissociation channels Li + CH and Li$^+$ + CH$^-$, the LiCH was found to be stable by approximately 77 kcal/mol.

\end{abstract}


\section{Introduction}\label{introduction}

\begin{figure}
\includegraphics[scale=0.25]{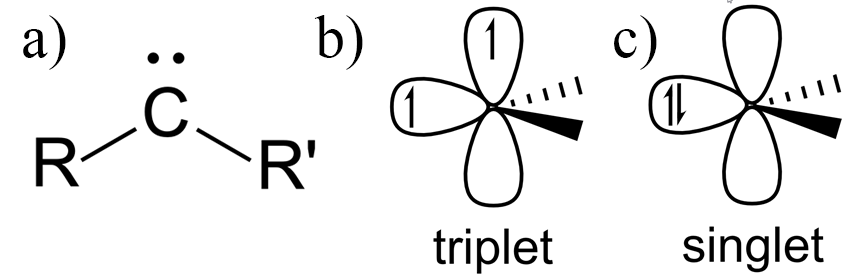} %
\caption{\label{fig:carbene} a) General carbene structure, a neutral divalent carbon with two unshared electrons. b) Diradical, triplet carbene with unpaired electrons in sp$^2$ and 2p-orbital on carbon.c) Singlet carbene with spin-paired electrons in sp$^2$ orbital and unfilled 2p-orbital. }
\end{figure}

Carbenes are a class of free radical compounds consisting of a neutral divalent carbon and two unshared electrons (Fig.~\ref{fig:carbene}a).
They were first proposed as intermediates in reaction mechanisms for the basic hydrolysis of chloroform over 150 years ago,\cite{geuther,nef}  long before they were ever recognized experimentally\cite{staudinger} and when even the notion of free radicals was not yet accepted.   Today, of course, carbene chemistry is ubiquitous and represents an important component of modern synthetic chemistry\cite{fremont}.

Methylene, CH$_2$, the simplest carbene, has its own rich history as theorists and experimentalists attempted to elucidate its ground-state electronic structure and geometry.\cite{SCHAEFER1100}  The very earliest {\em ab initio} study of CH$_2$ predicted a triplet ground electronic state  with a bent equilibrium geometry\cite{foster}, a year before the first experimental evidence was published suggesting a linear triplet ground state.\cite{herzberg} Over the course of the following decade, theoretical investigations continued to predict a bent triplet ground state with bond angle estimates ranging from 129\degree\ to 140\degree,\cite{SCHAEFER1100} with the most accurate calculation estimating the bond angle to be 135\degree.\cite{doi:10.1021/ja00719a039,SCHAEFER1100} Not until 1983 was the experimental bond angle of CH$_2$ established at 134\degree.\cite{jensen}. Today, it is well understood that most simple, dialkyl carbene compounds are bent with a triplet ground state, in which one electron occupies an sp$^2$-like hybridized orbital and another with parallel spin occupies a 2p orbital, both on the carbenic carbon center (Fig.~\ref{fig:carbene}b).\cite{anslyn} On the other hand, carbene substituents can stabilize the spin-paired singlet state, in which both electrons are in the sp$^2$ orbital (Fig.~\ref{fig:carbene}c).

Carbenes are highly reactive and undergo addition and insertion reactions, and the resulting products depend strongly on the electronic state of the carbene.  Singlet carbenes undergo concerted reactions that preserve stereogenic centers, whereas triplet carbenes undergo stepwise reactions that give rise to racemic mixtures.   So with the role that theoretical predictions have played in the understanding of carbene chemistry in mind, we predict a novel compound that straddles traditional carbene and organometallic reagent definitions---lithium carbene, LiCH.  While LiCH as a divalent compound with two unshared electrons is a carbene, as an alkali metal bonded to a CH, it also resembles the simplest experimentally known organometallic compound, methyl lithium, LiCH$_3$.   Given the prominence carbenes and organometallic compounds play in synthetic chemistry, we aim to establish computationally the potential existence of LiCH and explore its electronic properties.

To capture the multireference correlation potentially associated with a singlet diradical, we use a pair of two-electron reduced density matrix (2-RDM) methods.  To calculate accurate energy gaps, we employ the parametric 2-RDM  (P2RDM) method, a single reference method capable of capturing multireference correlation associated with a diradical. To assess the degree of the diradical character of LiCH, we use the variational 2-RDM (V2RDM) method, a multireference approach capable of capturing strong electron correlation.  Calculation results reveal a stable, triplet ground state and a diradical singlet exited state, both with a linear geometry.

\section{Computational Methods}\label{methods}

The parametric 2-RDM (P2RDM) method\cite{mazziotti2,mazziotti3,foley,valentine},  in Maple's Quantum Chemistry Package (QCP),\cite{maple, MapleQCP} was used to investigate the electronic structure of LiCH. Geometry optimizations of both singlet (S) and triplet (T) states were performed in the set of increasingly large cc-pVDZ, cc-pVTZ, cc-pVQZ, and cc-pV5Z basis sets.  Resulting electronic energies were used to calculate the spin energy gap, $\Delta E = E(\mathrm{S}) - E(\mathrm{T})$.  A positive \deltaE\ therefore corresponds to a triplet ground state. Correlation energies, natural orbital occupation numbers, natural orbital electron densities, Mulliken charges and net dipole moments were used to compare the two electronic states.   Electronic  energies and normal-mode vibrational zero-point energies of two dissociation channels, LiH + C and Li$^+$ + CH$^-$, were calculated in the cc-pVTZ basis and used to ascertain the stability of LiCH.  For comparison, calculations of methylene (CH$_2$) and methyl lithium (LiCH$_3$) were also performed in the cc-pV5Z basis.

The variational 2-RDM (V2RDM) method\cite{mazziotti4,mazziotti5,schlimgen,montgomery}, a multireference active-space method also implemented using the QCP, was used to compute natural orbital occupation numbers to compare diradical characters of both LiCH and CH$_2$ in the singlet state.   A [$N_e,N_o$] = [10,16] and [8,12] active space, where $N_e$ is the number of correlated electrons and $N_o$ is the number of active-space orbitals, was used for LiCH and CH$_2$, respectively, in the cc-pV5Z basis.

\section{Results}\label{results}

Molecular geometries for singlet and triplet states were optimized using the P2RDM method with cc-pVDZ, cc-pVTZ, cc-pVQZ, and cc-pV5Z basis sets for LiCH and the cc-pV5Z basis set for CH$_2$.  Bond lengths, $r_e$,  and bond angles, $\theta$, are listed in Table~\ref{tab:P2RDM}. The geometry of LiCH, roughly independent of electronic state, was found to be linear with Li-C and C-H bond distances being roughly 1.87 \AA\ and 1.09 \AA, respectively. In contrast, the geometry of CH$_2$ varies with electronic state, with the triplet (singlet) state being bent at  $\theta = 133.08^\mathrm{o}$ ($102.50^\mathrm{o}$) and $r_e = 1.074$ \AA (1.111\AA). The calculated CH$_2$ geometries are in good agreement with experimental values.\cite{herzberg,jensen,petek,doi:10.1021/jp953150i}.   The Li-C bond is also shorter than the corresponding bond in methyl lithium, which was calculated to be 1.97 \AA\ using the same model chemistry.
\begin{table}
\caption{\label{tab:P2RDM} Electronic energies and optimized geometries.}
\begin{threeparttable}[t]
  \begin{tabular}{|cccccc|c|}
  \hline
     	LiCH&  & cc-pVDZ &  cc-pVTZ &  cc-pVQZ &  cc-pV5Z & CH$_2$ cc-pV5Z (exp\tnote{a}\ )   \\
     	\hline
	&	$r_e$ Li-C (\AA)	&	1.8652	&	1.8733	&	1.8704	&	1.8710 &	\\
	&	$r_e$ C-H (\AA)		&	1.1124	&	1.0921	&	1.0950	&	1.0940 & 1.0735 (1.078) 	\\
    	Triplet & $\theta$ (deg)  & 180.00 & 180.00 & 180.00& 180.00 & 133.08 (136 )\\
     	&E (a.u.)	&	-45.9189	&	-45.9832	&	-46.0129	&	-46.0325 & -39.1236 	\\
     	&\ \ \ E$_\mathrm{corr} $ (a.u.)	&	-0.1297	&	-0.1818	&	-0.2088	&	-0.2277	& -0.1882 \\
    	\hline
     	&	$r_e$ Li-C (\AA)		&	1.8649	&	1.8703	&	1.8642	&	1.8639 &	\\
	&	$r_e$ C-H (\AA)		&	1.1153	&	1.0899	&	1.0962	&	1.0948 & 1.1023 (1.111)	\\
     	Singlet & $\theta$ (deg) & 180.00 & 180.00 & 180.00 & 180.00& 102.50 (102.4)\\
	&E	(a.u.)&	-45.8838	&	-45.9546	&	-45.9868	&	-46.0073 &-39.1083 	\\
	&\ \ \ E$_\mathrm{corr}$	(a.u.)&	-0.1618	&	-0.2183	&	-0.2471	&	-0.2666 & -0.2123	\\	
	\hline
	Gap &	{\bf $\Delta$E} (a.u.) &	0.0351		&	0.0286	&	0.0261			&	0.0252	& 0.01527 \\
	\hline
\end{tabular}
\begin{tablenotes}
     \item[a]  Refs.~\citenum{herzberg, jensen, petek, doi:10.1021/jp953150i}
   \end{tablenotes}
    \end{threeparttable}%
\end{table}

Electronic energies, correlation energies, and the singlet-triplet energy gap, \deltaE, are also listed in Table~\ref{tab:P2RDM}.
%
%
%
%
As expected, the ground electronic state for LiCH is a triplet, consistent with Hund's first rule and carbenes in general, including CH$_2$.\cite{doi:10.1021/ja01598a087,jensen,doi:10.1021/jp953150i} The P2RDM method captures more electron correlation for both states for increasing basis-set size, thereby lowering the energy.  A similar decrease is seen in the energy gap, with $\Delta E = 0.0252$ a.u. (15.8 kcal/mol) for the cc-pV5Z basis.  While the correlation energies are converged only to $\approx$0.02 a.u (12-13 kcal/mol), the rate of convergence for both states is very similar, leading to a $\Delta E$ converged to better than 0.001 a.u. (less than 1 kcal/mol).
The calculated $\Delta E$ is larger for LiCH than for CH$_2$, for which $\Delta E = 0.01527$ a.u. (9.58 kcal/mol), which is in consistent with the previously reported calculated (9.025 kcal/mol)\cite{doi:10.1021/jp953150i}   and experimental (8.998 kcal/mol) \cite{jensen,doi:10.1021/jp953150i} values.

As an initial means to explore electron correlation in both electronic states, occupation numbers, $n$, of the active natural orbitals were calculated and are presented in Table ~\ref{tab:occupation-nums} for the highest-occupied natural orbital (HONO) and lowest-occupied natural orbital (LUNO).
\begin{table}[ht]
\caption{\label{tab:occupation-nums} Natural orbital occupation numbers for the HONO and LUNO using P2RDM.}
\smallskip
\smallskip
\begin{tabular}{|cc|cccc|c|}
\hline									
	LiCH &	& cc-pVDZ	&	cc-pVTZ	&	cc-pVQZ	&	cc-pV5Z & CH$_2$ (cc-pV5Z)\\
\hline	
triplet	& $n_\mathrm{HONO}$ &	{\bf 0.986982}	&	{\bf 0.985438}	&	{\bf 0.984827}	&	{\bf 0.984644}	& {\bf 0.990024} \\
	& $n_\mathrm{LUNO}$ &	{\bf 0.986982}	&	{\bf 0.985438}	&	{\bf 0.984827}	&	{\bf 0.984644}& {\bf 0.988505}	\\
	\hline									
singlet	&$n_\mathrm{HONO}$ &	{\bf 1.338862}	&	{\bf 1.429027}	&	{\bf 1.460636}	&	{\bf 1.474931}&{\bf 1.866223} \\
	&	$n_\mathrm{LUNO}$ & {\bf 0.614963}	&	{\bf 0.521736}	&	{\bf 0.488719}	&	{\bf 0.473898}&{\bf 	0.104606} 	\\
\hline
\end{tabular}
\end{table}
%
%
The triplet states for both LiCH and CH$_2$ show the HONO and LUNO being singly occupied, as expected, and very nearly converged by the cc-pVQZ level.  It is interesting to compare the convergence trends of the fractional occupation numbers with the correlation energies seen in Tab.~\ref{tab:P2RDM}. For the triplet state, occupation numbers are very nearly converged even with the cc-pVDZ basis, but the correlation energy increases steadily with larger basis-set sizes, indicating the P2RDM method is capturing more dynamic correlation.

It is also interesting to discover the partial diradical character of LiCH for the singlet state, with $n_\mathrm{HONO} = 1.47$ and $n_\mathrm{LUNO} = 0.47$.  This  is in contrast to CH$_2$, which shows a spin pairing with $n_\mathrm{HONO} = 1.87$ and $n_\mathrm{LUNO} = 0.10$.  The P2RDM method is a single reference method and can be limited in the extent to which it can describe multireference correlation. In order to determine the degree of diradical character in LiCH, we used the V2RDM method and a [10,16] active space in the cc-pV5Z basis, which revealed even more diradical character of $n_\mathrm{HONO} = 1.12$ and $n_\mathrm{LUNO} = 0.84$ for the HONO and LUNO, respectively (Table~\ref{tab:occupation-nums-V2RDM}).
\begin{table}[ht]
\caption{\label{tab:occupation-nums-V2RDM} Natural orbital occupation numbers for frontier orbitals of singlet LiCH and CH$_2$ using V2RDM.}
\smallskip
\smallskip
\begin{tabular}{|c|ccc|ccc|}
\hline									
	 &	& LiCH	&		&		&	 CH$_2$ &\\
\hline								
active space& [10,12]	& [10,14]	& [10,16]		&	[8,8]	&	[8,10] & [8,12] \\
\hline
$n_\mathrm{HONO}$	&	1.176879	&	1.149403	&	{\bf 1.123649}	&	1.984138	&	1.952894	&	{\bf 1.943767}	\\
$n_\mathrm{LUNO}$	&	0.787064	&	0.817634	&	{\bf 0.839283}	&	0.015306	&	0.048400	&	{\bf 0.051513}	\\
\hline
\end{tabular}
\end{table}
  For CH$_2$, a V2RDM calculation with a [8,12] gave rise to $n_\mathrm{HONO} =  1.94$ and $n_\mathrm{LUNO} = 0.05$, an almost complete spin-pairing.  This stark difference in the electronic structure of LiCH and CH$_2$  may give rise to important differences in their respective reactivities in the singlet state.

\begin{figure}
\includegraphics[scale=0.6]{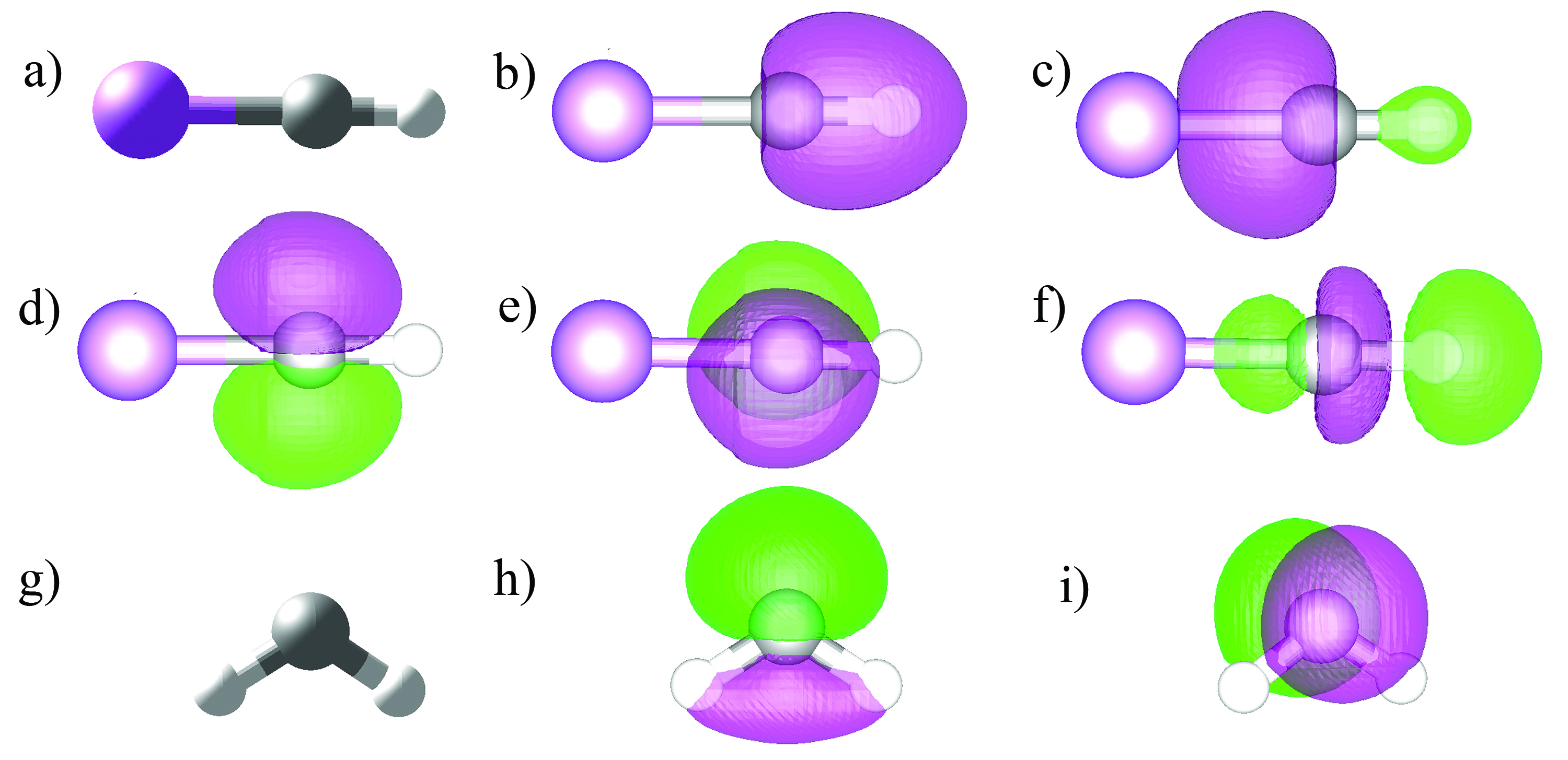} %
\caption{\label{fig:NOs} a) Optimized geometry for LiCH. b) - f) Natural orbitals for LiCH  for the b) HONO - 2, c) HONO - 1, d) HONO , e) LUNO, and f) LUNO+1.   g) Optimized geometry for CH$_2$.  h) - i) Natural orbitals for h) HONO and i) LUNO. }
\end{figure}

Natural orbital representations for singlet LiCH are provided in Fig.~\ref{fig:NOs}.
It is clear that the HONO - 2 (Fig.~\ref{fig:NOs} b)) is consistent with C-H bonding orbital, and HONO - 1 (Fig.~\ref{fig:NOs} c)) resembles a delocalized Li-C-H bond. The HONO and LUNO (Fig.~\ref{fig:NOs} d) and e)) appear to be non-bonding p-orbitals on the carbon. The LUNO+1 (Fig.~\ref{fig:NOs} f)) resembles a delocalized antibonding orbital. The HONO and LUNO for singlet CH$_2$ are also provided in (Fig.~\ref{fig:NOs}).  The overlap of the carbon 2p orbital with the two hydrogen s-orbitals lowers the energy of the HONO (Fig.~\ref{fig:NOs} h))  relative to the LUNO (Fig.~\ref{fig:NOs} i)), which corresponds to an unfilled p-orbital on the carbon.  The HONO is commonly described as an sp$^2$ hybridized orbital on the carbon.  One of the crucial differences between CH$_2$ and LiCH lies in their HONOs.  While the HONO of  CH$_2$ exhibits some sp$^2$ hydridization from bonding with the hydrogen atoms, leading to the molecule's signature bent geometry, the HONO of LiCH is a non-bonding p-orbital which contributes to its linear geometry and significant diradical character.

Table~\ref{tab:charges} contains Mulliken charges for each atom and the overall dipole moment for both singlet and triplet states for LiCH for cc-pV5Z.
\begin{table}[ht]
\caption{\label{tab:charges} Mulliken charges and dipole moment for LiCH}
\smallskip
\smallskip
\begin{tabular}{|c|cccc|}
\hline
	&	Li	&	C	&	H	&	Dipole	\\
\hline											
triplet	&	0.554376	&	-0.547623	&	-0.006753	&	-5.443756		\\
\hline	 		 		 		 		 		
singlet	&	0.561896	&	-0.567066	&	0.005171	&	-5.122367			\\
\hline
\end{tabular}
\end{table}
The charge delocalization on the Li and C is consistent with a polar covalent bond, also reflected in the delocalized HONO-1 bonding orbital (Fig.~\ref{fig:NOs} c)). The charge difference at 1.10 is slightly larger than for Li and C in LiCH$_3$, which is 1.00, calculated using P2RDM and cc-pVQZ basis.

As a final way to characterize the LiCH compound, we calculated the zero-point vibrational energies of triplet LiCH and compared the total energy to those of the most likely dissociation channels: LiH + C and Li$^+$ + CH$^-$. Electronic energies and vibrational frequencies for LiCH, LiH, and CH$^-$, and electronic energies for neutral C and Li$^+$ are provided in Table~\ref{tab:diss-channels}.
\begin{table}[ht]
\caption{\label{tab:diss-channels} Comparison of total energies for LiCH and  LiH + C  and Li$^+$ + CH$^-$ dissociation channels.}
\smallskip
\smallskip
\begin{tabular}{|c|c|}
\hline
E (LiCH) (a.u.)	& -45.9832	\\
$\tilde{\nu}_0$ (LiCH) (cm$^{-1}$)	&	3028.7098	\\
$\tilde{\nu}_1$ (LiCH) (cm$^{-1}$)	&	748.3092		\\
$\tilde{\nu}_2$, $\tilde{\nu}_3$ (LiCH) (cm$^{-1}$)	&	454.5938			\\
{\bf E (LiCH) + ZPE (a.u.)}	& -45.9726	\\
\hline
E$_\mathrm{elec}$ (LiH) (a.u.)	&	-8.0511		\\
$\tilde{\nu}_0$ (LiH) (cm$^{-1}$)	&	1720.1381			\\
E$_\mathrm{elec}$ (C) (a.u.)	&	-37.8194	\\
{\bf E$_\mathrm{channel} $ (LiH + C) (a.u.) }	&	-45.8494	\\
\hline
E$_\mathrm{elec}$ (CH$^-$) (a.u.)	&	-38.4415 	\\
$\tilde{\nu}_0$ (CH$^-$) (cm$^{-1}$)	&	2128.0730	 \\
E$_\mathrm{elec}$ (Li$^+)$ (a.u.)    	&	-7.2525  	\\
{\bf E$_\mathrm{channel}$ (Li$^+$ + CH$^-$)	 (a.u.)}&	-45.6892 	\\
\hline
\end{tabular}
\end{table}
The zero-point energy of LiCH is found to be lower than the LiH + C dissociation channel by approximately 77 kcal/mol and lower than the Li$^+$ + CH$^-$ channel by 178 kcal/mol, suggesting triplet LiCH is a stable gas-phase carbene-like compound.

\section{Discussion and Conclusions}\label{discussion}

In this work, using the parametric 2-RDM (P2RDM) and variational 2-RDM (V2RDM) methods and correlation-consistent basis sets from polarized double zeta (cc-pVDZ) through polarized 5-zeta  (cc-pV5Z)\cite{prascher2011a,dunning1989a}, we explore the electronic structure of LiCH, as yet undiscovered and potentially the simplest organometallic compound, and compare it to methylene, CH$_2$, the simplest carbene.   Like CH$_2$, LiCH is predicted to be a stable carbene compound with a triplet ground state and a singlet-triplet energy gap of 15.8 kcal/mol, compared to CH$_2$'s 9.58 kcal/mol calculated using the same model chemistry.   However, unlike CH$_2$, LiCH is expected to be linear with a diradical character in both the triplet ground state and the singlet excited state.

Organometallic complexes are critically important in synthetic chemistry, biology, and medicine.  Likewise, carbenes also have a crucial role in chemistry where as highly reactive species they promote important synthetic transformations.  In this computational study we have examine LiCH, a carbene that is potentially the simplest organometallic molecule.  While the LiCH carbene has connections to the conventional classes of carbenes---Fischer Schrock, and $N$-heterocyclic carbenes~\cite{fremont}, it is also fairly distinctive due to the involvement of only three atoms including a single metal atom.  Experimental isolation of LiCH should be possible based on the present computations and experimental results for related molecules such as the monohalocarbenes~\cite{nesbitt2018, nesbitt2020}.  Moreover, the simplicity of the carbene and possible derivatives may serve as a foundation for developing a broader class of carbenes with potential applications in chemistry, medicine, and materials.

\begin{acknowledgement}

This work was supported by the U.S. National Science Foundation Grant No. CHE-1565638 and the Chemical Sciences, Geosciences and Biosciences Division, Office of Basic Energy Sciences, Office of Science, U.S. Department of Energy, under Grant No. DE-FG02-92ER14305.

\end{acknowledgement}




\bibliography{refs}

\end{document}